\documentclass[manuscript]{emulateapj}

\shorttitle{New water fountain source IRAS 19190+1102}
\shortauthors{Day et al.}

\begin{document}

\title{Proper motions of H$_{2}$O masers in the water fountain source IRAS 19190+1102}

\author{F. M. Day and Y. M. Pihlstr\"om\altaffilmark{1}}
\affil{Department of Physics and Astronomy, University of New Mexico, 800 Yale Blvd NE, Albuquerque, NM 87131}
\email{fonda@unm.edu}

\author{M. J. Claussen}
\affil{National Radio Astronomy Observatory (NRAO), Array Operations Center, P.O. Box O, Socorro, NM 87801, USA}

\and

\author{R. Sahai}
\affil{Jet Propulsion Laboratory, MS183-900, California Institute of Technology, Pasadena, CA 91109, USA}

\altaffiltext{1}{Y. M. Pihlstr\"om is also an Adjunct Astronomer at the National Radio Astronomy Observatory.}

\begin{abstract}
We report on the results of two epochs of Very Long Baseline Array (VLBA) observations of the 22 GHz water masers toward IRAS 19190+1102.  The water maser emission from this object shows two main arc-shaped formations perpendicular to their NE-SW separation axis.  The arcs are separated by $\sim$280 mas in position, and are expanding outwards at an angular rate of 2.35 mas yr$^{-1}$.  We detect maser emission at velocities between -53.3 km s$^{-1}$ to +78.0 km s$^{-1}$ and there is a distinct velocity pattern where the NE masers are blueshifted and the SW masers are redshifted.  The outflow has a three-dimensional outflow velocity of 99.8 km s$^{-1}$ and a dynamical age of about 59 yr.  A group of blueshifted masers not located along the arcs shows a change in velocity of more than 25 km s$^{-1}$ between epochs, and may be indicative of the formation of a new lobe.  These observations show that IRAS 19190+1102 is a member of the class of ``water fountain'' pre-planetary nebulae displaying bipolar structure.
\end{abstract}

\keywords{masers -- stars: mass loss}

\section{Introduction}
 Intermediate mass stars (1-8 $M_{\odot}$) evolve from being asymptotic giant branch (AGB) stars into planetary nebulae (PNs) via a short transition phase during which the stars are classified as pre-planetary nebulae (PPNs).  AGB stars lose mass from a slow, dense wind with expansion velocities of 5--30 km s$^{-1}$ and exhibit roughly spherically symmetric circumstellar shells; however, PNs are often observed to have aspherical morphologies, including multipolar and elliptical structures (e.g., \citealt{sah98}; \citealt{sah00}; \citealt{sah07}).  Presumably it is during the PPN stage that some mechanism, responsible for the shaping of the wide variety of PN morphologies, becomes operational.  Due to the short lifetime of PPNs, few objects have been studied during this specific stage and the details of the evolution of an AGB star into a PN still remain unclear.

Understanding the short PPN stage is fundamental for understanding the final stages of intermediate mass stellar evolution.  As PPNs exhibit bipolar outflows (e.g., \citealt{sah07}), it is now believed that jets are important in shaping multipolar PNs \citep{sah98}.  Detailed observations of individual objects can provide information about the prevailing physical conditions under which the PN morphologies are formed, and about the progenitor star.  To study the effects of jets in these objects, kinematical information concerning the outflows is crucial.  Radio H$_{2}$O, OH, and SiO maser line emission can be used to aid classification of these objects in their late stages of evolution.  In ``water fountain'' nebulae, high-velocity H$_{2}$O masers (velocity spreads $>$ 50 km s$^{-1}$) are believed to trace high-velocity outflows.  Such high-velocity water masers were first discovered in \objectname{IRAS 16342-3814} \citep{lik88}, and several more water fountain PPN candidates have been discovered via their single dish spectra.  High angular resolution observations of water masers in IRAS 16342-3814, OH 12.8-0.9, IRAS 19134+2131, W43A, and IRAS 16552-3040 \citep{cla09,bob07,ima07,sua08} have confirmed their classification as water fountains, by showing a spatial and kinematical structure consistent with bipolar lobes.  Individual maser features persist on 1--3 year periods \citep{eng02} and can thus be used to trace dynamics of the gas.

Accurate distance measures to PPNs as well as PNs in our galaxy are sparse, thus limiting the accuracy of derived stellar properties.  Observations using the VLBA allow high precision ($<$ 1 mas at 22 GHz) astrometric studies of the maser features in PPNs, therefore enabling the possibility of performing trigonometric parallax measurements.  The measured proper motions of the masers is a combination of three components: motion of the gas relative to the central object, peculiar motion of the source within the galaxy, and parallax.  By measuring these motions we can determine distances to and dynamical ages of PPNs, thereby affording further understanding of the transition from the AGB to PN phase, including luminosity, mass, and mass loss rate.

\objectname{IRAS 19190+1102} is an OH and H$_{2}$O source with {\it Infrared Astronomical Satellite} ({\it IRAS}) colors characteristic of evolved stars ([25-60]$\lesssim$1.5 and [12-25]$\gtrsim$1.4, where [a-b]=2.5log($S_{b}/S_{a}$)).  \citet{lik89} first observed the OH \& H$_{2}$O emission in this source, finding the velocity range of the H$_{2}$O emission spanned more than 70 km s$^{-1}$, atypical of OH/IR stars.  In this paper, we present high spatial resolution data identifying IRAS 19190+1102 as a water fountain PPN.  We give initial results (two epochs) of observations of H$_{2}$O maser emission from IRAS 19190+1102, as well as a brief discussion of its far-infrared characteristics, and discuss its classification as a water fountain PPN.  This source is our first target to which we will perform parallax measurements on (to be presented in a subsequent paper).

\begin{figure}
\centering
\includegraphics[angle=270,scale=0.8]{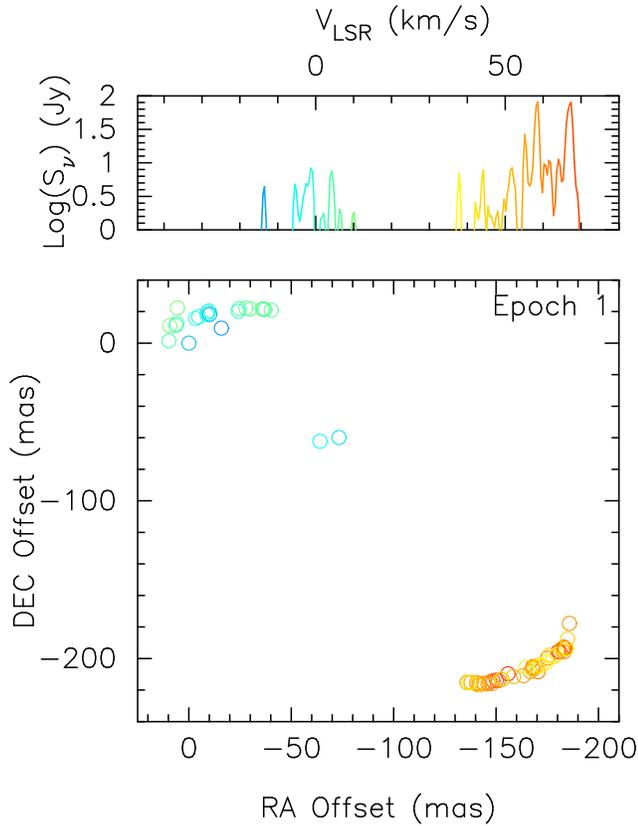}
\caption{2004 March 19 (Epoch 1)  Top panel: log plot of the spectral distribution of maser emission in IRAS 19190+1102.  Bottom panel:  spatial distribution of masers in IRAS 19190+1102, relative to the feature at (0,0).\label{fige1}}
\end{figure}

\section{Observations and data reduction} \label{obs}
VLBA spectral line observations of IRAS 19190+1102 were taken 2004 March 19 and 2006 May 31 of the 22.235 GHz 6$_{16}$ $\rightarrow$ 5$_{23}$ H$_{2}$O transition.  Observations were made using a 16 MHz band with dual circular polarization, consisting of 1024 channels centered around V$_{LSR}$=50 km s$^{-1}$.  Thus the velocity coverage was -58 to +157 km s$^{-1}$ with a spectral resolution of 0.21 km s$^{-1}$ per channel.  Total on-source time for each observation was $\sim$3.5 hr.  The data were correlated at the VLBA correlator in Socorro, NM.

The AIPS package was used to reduce and analyze the data.  The bandpass response and residual delays were determined using the source 3C454.3.  For these observations phase-referencing was not used so global fitting of the fringe rates were obtained by fringe fitting on a maser feature identified in both epochs.  Furthermore, an iterative self-calibration and imaging procedure was used to improve the SNR, hence absolute positions of the phase centers was lost.  Three of the ten antennas were flagged in both epochs due to poor SNR, and each epoch was imaged with a beam size of 1.58 mas$\times$0.96 mas.  The spectra were averaged to a velocity resolution of 0.42 km s$^{-1}$ per channel to improve the SNR for weaker masers.  Final cubes of both epochs had an rms of $\sigma\sim$10 mJy per channel for channels without line emission; rms noise in channels containing the brightest emission were found to be $\sigma$=20 mJy, 40 mJy for the first and second epochs, respectively.  

For a detection to be considered significant, emission had to span more than one 0.42 km s$^{-1}$ wide channel, and a cutoff flux density was set at 10 times the rms noise of each channel.  We obtained maser component parameters by fitting two-dimensional Gaussian functions to emission in individual channels.  Maser emission was spread over more than one channel, so we used a flux-density-squared, weighted-averaging scheme to calculate each maser feature's central velocity and position.  Errors from using the weighted-average were less than the velocity resolution, and less than 0.5 mas in position.

\begin{figure}
\centering
\includegraphics[angle=270,scale=0.8]{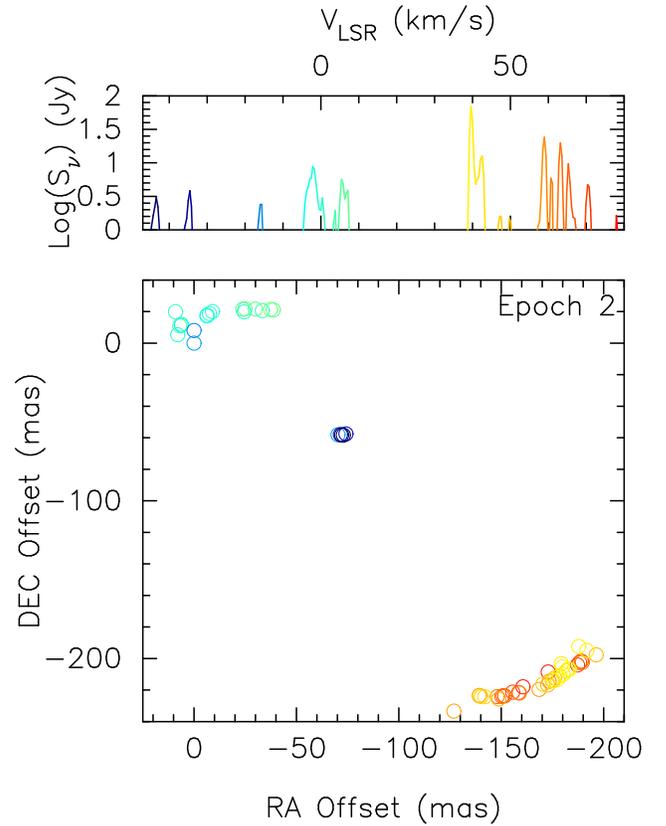}
\caption{2006 May 31 (Epoch 2)  Top panel: log plot of the spectral distribution of maser emission in IRAS 19190+1102.  Bottom panel:  spatial distribution of masers in IRAS 19190+1102, relative to the feature at (0,0).\label{fige2}}
\end{figure}

Because we were unable to obtain absolute positions of the maser features, we measured relative positions with respect to a feature being common in both epochs.  Although this feature drifted in velocity by almost 2 km s$^{-1}$ between the two epochs (indicating that perhaps some acceleration may be present which could be investigated with future epochs), its relative position with respect to several nearby masers is persistent and we therefore identify this maser as being the same in both epochs. Moreover, a test using a different reference maser did not change the results reported on in this paper.  Table ~\ref{tbl-1} (epoch 1) and Table~\ref{tbl-2} (epoch 2) include maser velocity, flux and its associated error, relative right ascension and its associated error, and relative declination and its error.  The flux assigned to a maser feature was the peak flux in the channels the feature spanned.  

\section{Results} \label{results}

The spatial distribution of maser emission in both epochs shows two main arc-shaped formations, arranged perpendicular to a NE-SW separation axis ($\sim$34 degrees east of north).  Spectral and spatial distributions of the H$_{2}$O masers are shown in Figures~\ref{fige1} (epoch 1) and ~\ref{fige2} (epoch 2).
Log plots were used to show the spectral distribution of the maser features in order to show more detail of the weaker, blueshifted features.  The northern features are blueshifted (-17.9 to 11.4 km s$^{-1}$), and southern features are redshifted (28.8 to 78.0 km s$^{-1}$).  Furthermore, both epochs have a third group of masers, which are not located within the arc-like structures, but instead appear along the axis of separation (see section~\ref{comparisons}).  The total velocity spread of the masers is $\Delta$V$\sim$90 km s$^{-1}$ in epoch 1 and $\Delta$V$\sim$130 km s$^{-1}$ in epoch 2.  

The arcs in epoch 1 are 53.65 mas (northern arc) and 62.44 mas (southern arc) across, whereas the arcs were only measured to be 49.23 mas (northern arc) and 62.62 mas (southern arc) across in epoch 2.  Even though the rms noise is slightly higher in the second epoch for the channels with bright emission, the masers in the edges of the arcs appear to have on average the same flux density as masers closer to the center and so we should have been able to detect these masers.  Therefore it appears as if the physical extent of conditions for masers to occur has decreased somewhat for the northern arc in the second epoch.  The average angular separation of the north and south masers is 269.27$\pm$0.24 mas in epoch 1 and 279.60$\pm$0.30 mas in epoch 2.  Using $d=$8.6 kpc as the distance to IRAS 19190+1102 (see section~\ref{dust}), the separation of the arcs is then 2316 AU and 2405 AU for epochs 1 and 2, respectively.

Since maser emission can be variable in both flux and velocity, and due to the large number of detections in each cube, it is difficult to analyze the kinematics of individual maser features.  Instead, we used the average angular separation to measure maser proper motions.  We assume that the northern and southern arcs are expanding at equal rates, and so we  halve the difference in angular separation to get an expansion rate of $\mu$=2.35$\pm$0.18 mas yr$^{-1}$.  The age for the jet, then, is estimated to be $\sim$59 yr, assuming that the proper motion is constant.  The average line of sight velocity of the the H$_{2}$O masers is $V_{r}=27.6$ km s$^{-1}$, which, when combined with the tangential speed of the masers ($V_{t}$=$\mu d$=95.9 km s$^{-1}$), yields a total 3D outflow speed, $V_{exp}=V_{r}^{2}+V_{t}^{2}$, of 99.8 km s$^{-1}$.  The inclination angle with respect to the plane of the sky was calculated to be $i=\tan^{-1}(V_{r}/V_{t})$=16.1$^{\circ}$.   By assuming that the dynamical center of the outflow is located at the midpoint along the axis of separation of the arc-like structures, we obtain jet opening angles of $\phi_{1} \sim$11$^{\circ}$--13$^{\circ}$ for epoch 1 and $\phi_{2}$ $\sim$10$^{\circ}$--13$^{\circ}$ for epoch 2.

\begin{figure}
\centering
\includegraphics[angle=270,scale=0.35]{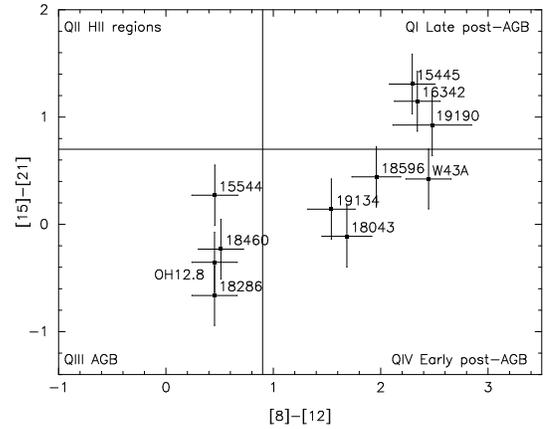}
\caption{{\it MSX} color-color diagram showing locations of known and candidate water fountain sources within the quadrants defined by \citet{sev02}.  The location of IRAS 19190+1102 strongly supports the conclusion that it truly is an evolved object, as opposed to a star forming region.\label{msx}}
\end{figure}

\section{Discussion}
\subsection{A True Water Fountain PPN} \label{wfppn}
In order for an object to be classified as a true water fountain PPN, it needs to display both water maser fountain characteristics as well as being an evolved star.  \citet{lik89} detected OH 1612, 1667 and H$_2$O masers in IRAS 19190+1102, but due to the OH and H$_{2}$O features' peculiar nature (e.g. a third peak in the 1612 MHz emission, H$_{2}$O emission spanning more than 70 km s$^{-1}$) and cold {\it IRAS} colors, it was not definitively classified as an evolved star at that time.  In {\it IRAS} color-color (e.g. [25-60] vs [12-25]) diagrams, there is some overlap in the regions occupied by HII regions, young stellar objects (YSOs), evolved stars and PNs.  Furthermore, \citet{mot07} include IRAS 19190+1102 (G046.2561-01.4763) as a candidate massive YSO based on its mid-infrared colors.  However, combining spectral properties of e.g. OH and H$_{2}$O with {\it IRAS} and  {\it Midcourse Space eXperiment} ({\it MSX}) colors can help distinguish evolved stars from very young objects (c.f. \citealt{sev02}).  

The location of IRAS 19190+1102 in the {\it IRAS} color-color diagram places it in the post-AGB region, and its {\it MSX} fluxes tentatively place it in quandrant I (late post-AGB region) as defined by \citet{sev02}.  Figure~\ref{msx} is a modification of the {\it MSX} color-color diagram of known and candidate water fountain sources made by \citet{sua08}, with IRAS 19190+1102 included as well as each source's three sigma error bars.  Although the 8$\mu$m and 12$\mu$m fluxes are at the limit of detection ({\it MSX} quality flag 1, with SNR$\approx$4.5 in both bands), the position of IRAS 19190+1102 lies well within the quadrant containing post-AGB objects.  Moreover, the 1612 MHz OH emission is stronger than the 1667 MHz OH emission \citep{lik89} which is also consistent with what is observed in the thick circumstellar shells of evolved OH/IR stars.  Thus we postulate that IRAS 19190+1102 truly is an evolved object.  Our VLBA observations confirm that the bipolar structure, outward expansion, and high-velocity H$_{2}$O masers of IRAS 19190+1102 make it a water fountain PPN.

\subsection{Molecular Gas and Dust} \label{dust}
\begin{figure}
\centering
\includegraphics[angle=270,scale=0.35]{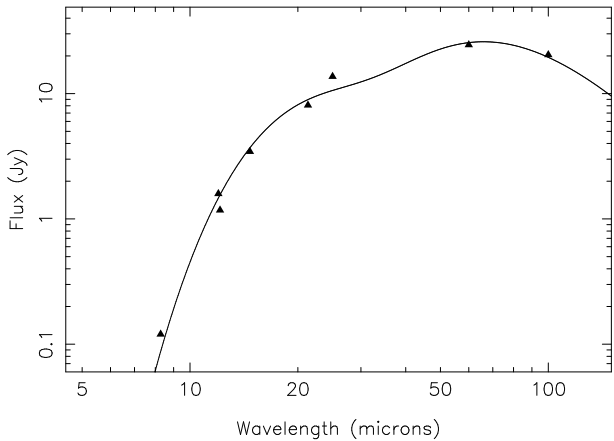}
\caption{The far-infrared emission observed from IRAS 19190+1102.  Also shown is the calculated flux (continuous line) with dust emissivity $\propto \nu^{1.5}$ for a particular best-fit model.  At least two different dust components, with different temperatures and masses, are required to fit the data.\label{sed}}
\end{figure}

The infrared emission in IRAS 19190+1102 can be used as an estimate of the dust mass, and its temperature \citep{sah91}.  Using the {\it IRAS} fluxes of IRAS 19190+1102 (1.59, 13.67, 24.52, and 20.41 Jy in the 12, 25, 60 and 100$\mu$ bands) as well as its {\it MSX} fluxes (0.12017, 1.17535, 3.4499, and 8.0921 Jy in the 8, 12, 15, and 21$\mu$ bands), we find that two dust components with temperatures of 46 K and 120 K are required to fit the data according to the model described in \citet{sah91}.  The masses of the two dust components are, respectively, $M_{dust}=0.7\times10^{-3}(d/1$ kpc)$^{2}$ $M_{\odot}$ and $0.4\times 10^{-5}(d/1$ kpc)$^{2}$ $M_{\odot}$.  Figure~\ref{sed} shows the results of the fit of the data with a peak of the spectral energy distribution (SED) between 60 and 100$\mu$m.  

The distance to IRAS 19190+1102 is poorly known, and has been quoted as low as 1.5 kpc \citep{pre88}.  \citet{pre88} used the modified Shklovsky method to determine the distance to IRAS 19190+1102, whereby the ionized mass of a PN is related to its electron density .  However, \citet{van95} did not detect 6-cm continuum emission in IRAS 19190+1102 down to 3 to 4 mJy, so most of its nebular mass is not ionized, making a distance of 1.5 kpc suspect.  We instead estimate the distance assuming that IRAS 19190+1102 has a typical PPN luminosity of 6000 L$_{\odot}$\citep{sah07}, in which case it can be placed as far away as $d=8.6$ kpc using our SED fit of the far-infrared fluxes.  In this paper we therefore adopted a distance to IRAS 19190+1102 of 8.6 kpc. 

Furthermore, by assuming a distance of $d=8.6$ kpc and a typical gas-to-dust ratio of 200, the total dust mass of IRAS 19190+1102 is 0.05 $M_{\odot}$ and the total gas mass is close to 10 $M_{\odot}$.  The mass estimate is, of course, very uncertain; the largest uncertainty is due to the mass derivation's square dependence on distance, followed by uncertainties in the dust mass derivation and gas-to-dust ratio.  Hence our estimate of 10 $M_{\odot}$ is not necessarily inconsistent with IRAS 19190+1102 being a PPN.  We merely emphasize that the mass for this object is likely to be large--in the upper end of the PPN progenitor stellar mass range.

\subsection{Comparisons to Other Water Fountain Sources} \label{comparisons}

There exist at least five other known water fountain sources: IRAS 16342-3814, \objectname{OH 12.8-0.9}, \objectname{IRAS 19134+2131}, \objectname{W43A}, and \objectname{IRAS 16552-3050} \citep{cla09,bob07,ima07,sua08} as well as numerous other candidates (cf. \citealt{imb07} and references therein).  The age we measure for IRAS 19190+1102 of $\sim$59 years is similar to ages of other water fountain nebulae, which lie in the range of 35--130 years.  The linear separation of the arcs in IRAS 19190+1102 of 2405 AU at 8.6 kpc places it within the range for other water fountain objects (880 AU for OH 12.8-0.9 up to 6000 AU for IRAS 16342-3814).  Our detection of highly collimated outflows ($\phi \sim$10$^{\circ}$--13$^{\circ}$) in this source is characteristic of the other water fountain pre-planetary nebulae, which have $\phi <$15$^{\circ}$ (with the exception of IRAS 16552-3050, in which the opening angle of the outflow is rather large, 70$^{\circ}$ \citep{sua08}).

\begin{figure}
\centering
\includegraphics[angle=270,scale=0.35]{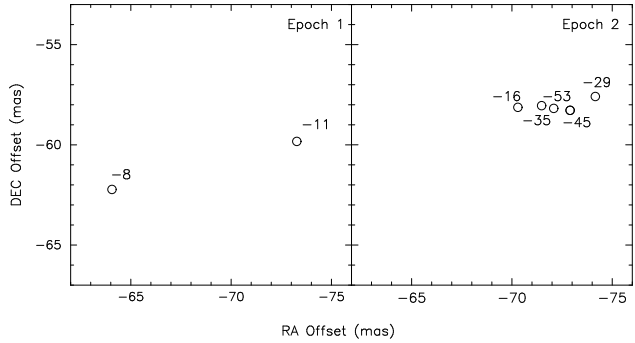}
\caption{A zoom in of the centrally located masers in epoch 1 (left) and epoch 2 (right).  The numbers represent the LSR velocity of each maser.  The masers in epoch 2 have no counterpart to those in epoch 1, and their velocities are also much different, suggesting they may be the signposts of a new lobe structure which is in an early, unstable formation phase.\label{blue}}
\end{figure}

As in the case of IRAS 16342-3814, there is no apparent systematic change in velocity along the arcs and there is a group of blue-shifted masers not distributed along the arcs (epoch 1: -11 to -8 km s$^{-1}$; epoch 2: -16.1 to -53.3 km s$^{-1}$; see Figure~\ref{blue}).  Interestingly, in contrast to IRAS 16342-3814, the most blueshifted features in IRAS 19190+1102 in epoch 2 correspond to the group not along the arcs, but are located close to the dynamical center, at a radial distance of about 335--385 AU from the latter.  Although this location depends on the unknown distance, it seems unlikely that these masers could be remnants from the AGB OH/IR phase, where H$_{2}$O masers occur between 10--100 AU from the star.

The average velocity of the masers in this central region has changed by more than 25 km s$^{-1}$ between epochs 1 and 2.  More observations are necessary to understand their exact nature, and we can just speculate that the masers were created from an episodic or precessing jet hitting new material.  Perhaps the dramatic change in velocities of these features could be a result of a new jet hitting clumps with different densities and accelerating them to different speeds.  It is also possible that this jet outflow has just started operating and consequently has not reached a steady state like the bipolar jet responsible for creating the slowly expanding arc-like structures.  These features may indicate the earliest stages of the formation of a new lobe, and so it is possible that IRAS 19190+1102 will eventually develop minor lobes (c.f. the PPNs Frosty Leo \citep{saa00} and AFGL 2688 \citep{saa98}) or become a multipolar nebula (c.f. IRAS 19475+3119, IRAS 19024+0044 \citep{sah07}).  Thus, monitoring the H$_{2}$O masers in IRAS 19190+1102 should be given high priority in future observational studies of water fountain pre-planetary nebulae, for understanding the earliest stages of the formation of bipolar and possibly multipolar structure in these objects.

\section{Conclusions}
Using the VLBA, we have observed the H$_{2}$O masers toward IRAS 19190+1102 with very high angular resolution; these observations indicate that the high-velocity masers span a velocity range $\Delta$V$>$130 km s$^{-1}$ and lie in a bipolar structure.  The afore mentioned properties indicate that IRAS 19190+1102 is one of six known ``water fountain'' sources, i.e. very young pre-planetary nebulae in which the presence of collimated, high-velocity outflows is manifest.  The arclike structures appear to lie perpendicular to a NE-SW axis, with the blueshifted features to the north and redshifted features to the south.  The arclike distributions are suggestive of bow shocks produced by a collimated jet colliding with surrounding material.  The proper motions of the H$_{2}$O masers suggest that the dynamical age of the jet is $\sim$59 yr.  Additionally, a group of blueshifted masers not located along the arcs was detected in both epochs.  These ``central" masers showed a dramatic change ($>$25 km s$^{-1}$) in velocity, which we suggest may signal the beginning of a new lobe due to an episodic or precessing jet.

We have obtained five additional epochs of IRAS 19190+1102 using the VLBA, using the  22 GHz line of H$_{2}$O and the 1612, 1665, 1667, 1720 MHz OH lines.  With these observations we anticipate establishing an accurate estimate to the distance of IRAS 19190+1102 via parallax measurements, and thus more precise estimates of its stellar properties.

\acknowledgments

The National Radio Astronomy Observatory is a facility of the National Science Foundation operated under cooperative agreement by Associated Universities, Inc.
RS thanks NASA for partially funding this work via LTSA award (\# NM0710651/399-20-40-06), ADP award (\# NM0710651/399-20-00-08) and HST/GO award (\# GO-09801.01-A).

{\it Facilities:} \facility{VLBA}

\clearpage
\begin{deluxetable}{lcccccr|lcccccr}
\tabletypesize{\footnotesize}
\tablecolumns{14}
\tablecaption{Maser features in epoch 1 (2004 March 19).  Positions are relative to the feature with position (0,0) (in boldface).\label{tbl-1}}
\tablewidth{0pt}
\tablehead{
\multicolumn{7}{c}{Blueshifted Masers}& \multicolumn{7}{c}{Redshifted Masers}\\
\colhead{$v_{LSR}$} & \colhead{$S_{\nu}$} & \colhead{$\sigma_{S_{\nu}}$} & \colhead{$\Delta$RA} & \colhead{$\sigma_{RA}$} & \colhead{$\Delta$DEC} & \colhead{$\sigma_{DEC}$} & \colhead{$v_{LSR}$} & \colhead{$S_{\nu}$} & \colhead{$\sigma_{S_{\nu}}$} & \colhead{$\Delta$RA} & \colhead{$\sigma_{RA}$} & \colhead{$\Delta$DEC} & \colhead{$\sigma_{DEC}$}\\
\colhead{{\scriptsize km/s}} & \colhead{{\scriptsize Jy/beam}} & \colhead{{\scriptsize Jy/beam}} & \colhead{{\scriptsize mas}} & \colhead{{\scriptsize mas}} & \colhead{{\scriptsize mas}} & \colhead{{\scriptsize mas}}& \colhead{{\scriptsize km/s}} & \colhead{{\scriptsize Jy/beam}} & \colhead{{\scriptsize Jy/beam}} & \colhead{{\scriptsize mas}} & \colhead{{\scriptsize mas}} & \colhead{{\scriptsize mas}} & \colhead{{\scriptsize mas}}
}
\startdata
-17.30 & 0.269 & 0.008 & -15.878 & 0.021 & 9.522   & 0.010 & 28.77 & 0.306  & 0.008 & -164.606 & 0.015 & -205.541 & 0.004\\
-14.00 & 0.721 & 0.008 & -10.208 & 0.012 & 18.288  & 0.020 & 37.83 & 5.430  & 0.012 & -174.451 & 0.016 & -202.325 & 0.012\\
{\bf -13.54} & {\bf 4.161} & {\bf 0.007} &  {\bf 0.000}  & {\bf 0.018} & {\bf 0.000}  & {\bf 0.024} & 37.84 & 1.214  & 0.012 & -152.751 & 0.007 & -212.977 & 0.032\\
-10.68 & 0.534 & 0.008 & -73.276 & 0.003 & -59.827 & 0.008 & 40.59 & 0.231  & 0.008 & -172.031 & 0.070 & -204.633 & 0.132\\
-8.18  & 0.439 & 0.007 & -64.060 & 0.009 & -62.226 & 0.005 & 41.27 & 0.370  & 0.008 & -184.379 & 0.011 & -193.792 & 0.028\\
-7.56  & 0.369 & 0.008 & -9.003  & 0.028 & 18.070  & 0.123 & 42.14 & 2.440  & 0.009 & -179.817 & 0.002 & -197.305 & 0.017\\
-5.44  & 0.813 & 0.010 & -4.817  & 0.030 & 16.683  & 0.017 & 42.72 & 0.632  & 0.009 & -137.225 & 0.199 & -215.326 & 0.029\\
-5.38  & 3.674 & 0.010 & -9.117  & 0.013 & 19.600  & 0.018 & 43.05 & 0.454  & 0.009 & -177.804 & 0.071 & -198.968 & 0.068\\
-2.09  & 3.076 & 0.010 & -24.094 & 0.258 & 20.225  & 0.178 & 44.07 & 7.074  & 0.013 & -175.714 & 0.064 & -199.634 & 0.054\\
-2.06  & 1.105 & 0.010 & -9.897  & 0.210 & 20.453  & 0.082 & 45.50 & 1.678  & 0.009 & -140.549 & 0.039 & -216.692 & 0.009\\
1.32   & 1.050 & 0.009 & 5.869   & 0.140 & 12.163  & 0.103 & 45.56 & 0.732  & 0.009 & -175.942 & 0.018 & -197.951 & 0.004\\
1.97   & 0.625 & 0.009 & 9.788   & 0.015 & 1.404   & 0.024 & 46.41 & 0.470  & 0.009 & -169.639 & 0.023 & -205.574 & 0.022\\
2.34   & 0.763 & 0.009 & 6.391   & 0.028 & 11.207  & 0.099 & 47.05 & 0.244  & 0.009 & -184.880 & 0.003 & -187.510 & 0.035\\
3.74   & 0.658 & 0.010 & -37.198 & 0.015 & 21.376  & 0.011 & 47.95 & 0.631  & 0.009 & -170.399 & 0.062 & -204.368 & 0.089\\
4.17   & 5.537 & 0.011 & -24.565 & 0.215 & 21.907  & 0.058 & 48.11 & 0.484  & 0.009 & -181.991 & 0.096 & -193.642 & 0.085\\
4.84   & 0.244 & 0.009 & -40.205 & 0.065 & 20.878  & 0.015 & 48.83 & 2.220  & 0.009 & -135.934 & 0.403 & -215.366 & 0.168\\
5.24   & 0.148 & 0.008 & -35.861 & 0.065 & 21.729  & 0.019 & 51.30 & 2.189  & 0.011 & -168.293 & 0.033 & -206.623 & 0.044\\
6.20   & 1.704 & 0.008 & -29.838 & 0.056 & 21.804  & 0.029 & 51.34 & 4.629  & 0.012 & -158.474 & 0.045 & -211.435 & 0.070\\
6.71   & 1.126 & 0.008 & -27.820 & 0.011 & 22.272  & 0.009 & 52.34 & 3.004  & 0.011 & -135.752 & 0.031 & -215.029 & 0.013\\
7.98   & 0.636 & 0.008 & -36.420 & 0.042 & 21.617  & 0.024 & 54.76 & 0.315  & 0.014 & -167.874 & 0.022 & -206.543 & 0.019\\
9.93   & 1.902 & 0.008 & 5.509   & 0.016 & 22.362  & 0.014 & 55.31 & 22.291 & 0.025 & -144.238 & 0.060 & -216.010 & 0.142\\
11.39  & 0.358 & 0.008 & 9.311   & 0.016 & 10.988  & 0.020 & 55.61 & 0.595  & 0.017 & -141.600 & 0.037 & -216.697 & 0.196\\
 & & & & & & & 55.70 & 1.347  & 0.016 & -137.836 & 0.059 & -215.194 & 0.012\\
 & & & & & & & 55.94 & 1.398  & 0.017 & -166.764 & 0.027 & -208.344 & 0.015\\
 & & & & & & & 56.51 & 1.092  & 0.010 & -140.474 & 0.022 & -215.353 & 0.015\\
 & & & & & & & 57.15 & 0.255  & 0.012 & -168.077 & 0.009 & -205.775 & 0.075\\
 & & & & & & & 57.71 & 3.272  & 0.024 & -147.948 & 0.025 & -215.954 & 0.007\\
 & & & & & & & 58.38 & 65.292 & 0.025 & -163.341 & 0.025 & -210.903 & 0.006\\
 & & & & & & & 58.86 & 1.089  & 0.025 & -167.834 & 0.400 & -206.499 & 0.315\\
 & & & & & & & 59.96 & 0.190  & 0.010 & -185.803 & 0.020 & -177.787 & 0.003\\
 & & & & & & & 60.30 & 7.218  & 0.014 & -168.010 & 0.005 & -205.339 & 0.011\\
 & & & & & & & 60.51 & 0.398  & 0.012 & -146.040 & 0.009 & -215.964 & 0.012\\
 & & & & & & & 60.62 & 1.412  & 0.012 & -183.719 & 0.017 & -193.356 & 0.006\\
 & & & & & & & 61.33 & 2.252  & 0.014 & -141.725 & 0.017 & -216.260 & 0.039\\
 & & & & & & & 61.56 & 1.366  & 0.014 & -183.073 & 0.005 & -195.755 & 0.004\\
 & & & & & & & 61.73 & 7.104  & 0.014 & -143.810 & 0.012 & -215.744 & 0.012\\
 & & & & & & & 63.79 & 4.307  & 0.012 & -175.093 & 0.011 & -199.731 & 0.029\\
 & & & & & & & 64.22 & 1.780  & 0.012 & -170.420 & 0.017 & -208.431 & 0.013\\
 & & & & & & & 64.24 & 5.242  & 0.012 & -152.787 & 0.173 & -213.286 & 0.051\\
 & & & & & & & 65.07 & 0.523  & 0.011 & -151.155 & 0.043 & -213.977 & 0.084\\
 & & & & & & & 65.07 & 0.264  & 0.011 & -167.891 & 0.009 & -205.154 & 0.011\\
 & & & & & & & 65.65 & 4.106  & 0.014 & -183.197 & 0.035 & -192.793 & 0.090\\
 & & & & & & & 65.84 & 4.726  & 0.014 & -146.895 & 0.011 & -214.922 & 0.043\\
 & & & & & & & 67.10 & 68.043 & 0.026 & -182.992 & 0.052 & -193.416 & 0.079\\
 & & & & & & & 68.01 & 3.221  & 0.026 & -150.133 & 0.020 & -213.702 & 0.013\\
 & & & & & & & 68.75 & 2.163  & 0.010 & -148.472 & 0.033 & -214.135 & 0.103\\
 & & & & & & & 71.67 & 0.289  & 0.008 & -181.106 & 0.383 & -195.699 & 0.081\\
 & & & & & & & 73.68 & 0.306  & 0.008 & -180.275 & 0.041 & -195.677 & 0.011\\
 & & & & & & & 74.22 & 0.400  & 0.008 & -155.831 & 0.030 & -209.769 & 0.019\\
\enddata
\end{deluxetable}

\clearpage
\begin{deluxetable}{lcccccr|lcccccr}
\tabletypesize{\footnotesize}
\tablecolumns{14}
\tablecaption{Maser features in epoch 2 (2006 May 31).  Positions are relative to the feature with position (0,0) (in boldface).\label{tbl-2}}
\tablewidth{0pt}
\tablehead{
\multicolumn{7}{c}{Blueshifted Masers}& \multicolumn{7}{c}{Redshifted Masers}\\
\colhead{$v_{LSR}$} & \colhead{$S_{\nu}$} & \colhead{$\sigma_{S_{\nu}}$} & \colhead{$\Delta$RA} & \colhead{$\sigma_{RA}$} & \colhead{$\Delta$DEC} & \colhead{$\sigma_{DEC}$} & \colhead{$v_{LSR}$} & \colhead{$S_{\nu}$} & \colhead{$\sigma_{S_{\nu}}$} & \colhead{$\Delta$RA} & \colhead{$\sigma_{RA}$} & \colhead{$\Delta$DEC} & \colhead{$\sigma_{DEC}$}\\
\colhead{{\scriptsize km/s}} & \colhead{{\scriptsize Jy/beam}} & \colhead{{\scriptsize Jy/beam}} & \colhead{{\scriptsize mas}} & \colhead{{\scriptsize mas}} & \colhead{{\scriptsize mas}} & \colhead{{\scriptsize mas}}& \colhead{{\scriptsize km/s}} & \colhead{{\scriptsize Jy/beam}} & \colhead{{\scriptsize Jy/beam}} & \colhead{{\scriptsize mas}} & \colhead{{\scriptsize mas}} & \colhead{{\scriptsize mas}} & \colhead{{\scriptsize mas}}
}
\startdata
-53.27 & 0.287  & 0.016 & -72.900  & 0.005 & -58.278  & 0.041 & 37.66  & 0.611  & 0.010 & -183.179 & 0.007 & -206.927 & 0.003\\
-44.68 & 2.124  & 0.018 & -72.083  & 0.111 & -58.181  & 0.067 & 39.49  & 1.106  & 0.041 & -179.300 & 0.017 & -203.316 & 0.005\\
-34.90 & 2.538  & 0.018 & -71.480  & 0.012 & -58.045  & 0.008 & 39.57  & 0.358  & 0.017 & -191.745 & 0.043 & -194.892 & 0.026\\
-29.69 & 0.481  & 0.009 & -74.157  & 0.005 & -57.590  & 0.004 & 39.59  & 1.044  & 0.041 & -179.791 & 0.035 & -205.364 & 0.409\\
-17.93 & 0.308  & 0.008 & -0.009   & 0.014 & 8.000    & 0.011 & 39.59  & 1.744  & 0.041 & -182.769 & 0.030 & -206.719 & 0.038\\
-16.06 & 0.574  & 0.011 & -70.295  & 0.013 & -58.129  & 0.011 & 39.63  & 2.110  & 0.405 & -177.445 & 0.029 & -213.267 & 0.019\\
{\bf -15.94} & {\bf 2.441}  & {\bf 0.011} & {\bf 0.000}    & {\bf 0.012} & {\bf 0.000}    & {\bf 0.018} & 39.63  & 27.595 & 0.040 & -180.170 & 0.008 & -210.045 & 0.016\\
-8.18  & 0.762  & 0.008 & 6.199    & 0.003 & 11.315   & 0.004 & 39.66  & 1.736  & 0.041 & -187.803 & 0.029 & -192.371 & 0.053\\
-7.05  & 0.411  & 0.008 & -6.247   & 0.025 & 17.375   & 0.010 & 39.83  & 1.065  & 0.043 & -175.564 & 0.041 & -214.059 & 0.091\\
-2.75  & 3.761  & 0.025 & -24.490  & 0.061 & 19.822   & 0.124 & 40.36  & 10.918 & 0.043 & -178.400 & 0.147 & -211.032 & 0.237\\
-2.40  & 0.562  & 0.013 & 7.966    & 0.026 & 5.293    & 0.032 & 42.71  & 1.455  & 0.032 & -181.794 & 0.064 & -207.896 & 0.059\\
-2.07  & 1.541  & 0.025 & -6.409   & 0.011 & 17.754   & 0.029 & 44.21  & 0.799  & 0.011 & -141.922 & 0.116 & -224.247 & 0.089\\
-1.79  & 2.538  & 0.020 & -9.064   & 0.200 & 20.002   & 0.075 & 46.21  & 0.868  & 0.012 & -170.364 & 0.005 & -216.287 & 0.008\\
-0.57  & 1.429  & 0.015 & -7.550   & 0.007 & 18.766   & 0.006 & 47.30  & 1.107  & 0.014 & -140.195 & 0.004 & -224.090 & 0.009\\
0.04   & 0.461  & 0.012 & 9.036    & 0.016 & 19.963   & 0.017 & 48.33  & 0.553  & 0.011 & -175.995 & 0.031 & -212.470 & 0.110\\
0.42   & 1.904  & 0.016 & -33.355  & 0.114 & 20.650   & 0.017 & 49.81  & 0.658  & 0.013 & -139.338 & 0.003 & -223.616 & 0.004\\
1.48   & 0.810  & 0.010 & 6.161    & 0.125 & 12.045   & 0.035 & 49.89  & 1.039  & 0.012 & -174.813 & 0.037 & -213.856 & 0.062\\
3.61   & 1.680  & 0.015 & -23.813  & 0.010 & 21.549   & 0.009 & 51.55  & 0.594  & 0.012 & -172.694 & 0.022 & -216.830 & 0.019\\
4.49   & 0.525  & 0.012 & -38.643  & 0.027 & 21.157   & 0.049 & 51.69  & 0.390  & 0.012 & -139.226 & 0.079 & -223.407 & 0.028\\
5.46   & 3.033  & 0.022 & 6.980    & 0.068 & 11.041   & 0.151 & 51.69  & 0.569  & 0.012 & -196.323 & 0.013 & -197.690 & 0.014\\
6.52   & 1.812  & 0.016 & -24.885  & 0.022 & 21.754   & 0.029 & 53.88  & 0.534  & 0.010 & -148.819 & 0.042 & -225.731 & 0.159\\
7.02   & 1.485  & 0.016 & -29.909  & 0.005 & 21.719   & 0.009 & 55.79  & 0.544  & 0.013 & -126.853 & 0.027 & -233.320 & 0.008\\
9.35   & 0.516  & 0.009 & -37.431  & 0.010 & 21.424   & 0.013 & 57.41  & 0.871  & 0.016 & -168.475 & 0.037 & -219.583 & 0.012\\
 & & & & & & & 58.91  & 8.976  & 0.038 & -173.337 & 0.007 & -214.489 & 0.006\\
 & & & & & & & 60.82  & 2.909  & 0.052 & -158.830 & 0.005 & -221.978 & 0.064\\
 & & & & & & & 63.20  & 8.535  & 0.039 & -148.011 & 0.013 & -224.128 & 0.015\\
 & & & & & & & 64.13  & 0.418  & 0.016 & -189.618 & 0.033 & -202.385 & 0.005\\
 & & & & & & & 65.27  & 4.154  & 0.031 & -150.254 & 0.024 & -224.042 & 0.024\\
 & & & & & & & 65.77  & 1.353  & 0.026 & -151.707 & 0.010 & -223.699 & 0.015\\
 & & & & & & & 66.77  & 1.214  & 0.015 & -189.064 & 0.056 & -201.638 & 0.064\\
 & & & & & & & 67.52  & 0.546  & 0.012 & -150.918 & 0.081 & -223.606 & 0.020\\
 & & & & & & & 68.50  & 0.526  & 0.010 & -158.452 & 0.004 & -221.553 & 0.008\\
 & & & & & & & 70.42  & 2.363  & 0.022 & -188.246 & 0.136 & -202.452 & 0.075\\
 & & & & & & & 70.94  & 0.763  & 0.022 & -155.552 & 0.009 & -221.343 & 0.014\\
 & & & & & & & 72.37  & 0.983  & 0.013 & -187.238 & 0.007 & -204.143 & 0.025\\
 & & & & & & & 76.14  & 0.648  & 0.010 & -160.555 & 0.006 & -217.938 & 0.012\\
 & & & & & & & 78.04  & 1.391  & 0.015 & -172.903 & 0.009 & -208.518 & 0.014\\
\enddata
\end{deluxetable}


\begin{thebibliography}{}
\bibitem[Boboltz \& Marvel(2007)]{bob07} Boboltz, D.A., \& Marvel, K.B.  2007, \apj, 665, 680
\bibitem[Claussen et al.(2009)]{cla09} Claussen, M.J., Sahai, R., \& Morris, M.R.  2009, \apj, 29, 219
\bibitem[Engels(2002)]{eng02} Engels, D.  2002, \aap, 388, 252
\bibitem[Imai et al.(2007)]{ima07} Imai, H., Sahai, R., \& Morris, M.  2007, \apj, 669, 424
\bibitem[Imai(2007)]{imb07} Imai, H.  2007, in IAU Symp. 242, Astrophysical Masers and Their Environments, ed. W. Baan \& J. Chapman (Cambridge: Cambridge Univ. Press), 279
%\bibitem[Kwok(1993)]{kwo93} Kwok, S.  1993, ARAA, 31, 63
%\bibitem[Kwok et al.(1996)]{kwo96} Kwok, S., Hrivnak, B.J., Zhang, C.Y., \& Langill, P.L.  1996, \apj, 472, 287
\bibitem[Likkel(1989)]{lik89} Likkel, L.  1989, \apj, 344, 350
\bibitem[Likkel \& Morris(1988)]{lik88} Likkel, L., \& Morris, M.  1988, \apj, 329, 914
%\bibitem[Likkel et al.(1992)]{lik92} Likkel, L., Morris, M., \& Maddalena, R.J.  1992, A\&A, 256, 581
\bibitem[Mottram et al.(2007)]{mot07} Mottram, J.C., Hoare, M.G., Lumsden, S.L., Oudmaijer, R.D., Urquhart, J.S., Sheret, T.L., Clarke, A.J., \& Allsopp, J.  2007, \aap, 476, 1019
%\bibitem[Ney et al.(1975)]{ney75} Ney, E.P., Merrill, K.M., Becklin, E.E., Neugebauer, G., \& Wynn-Williams, C.G.  1975, ApJ, 198, L129
%\bibitem[Nyman et al.(1998)]{nym98} Nyman, L.-A., Hall, P.J., \& Olofsson, H.  1998, \aaps, 127, 185
\bibitem[Preite-Martinez(1988)]{pre88} Preite-Martinez, A.  1988, \aap, 76, 317
\bibitem[Sahai et al.(2000a)]{saa00} Sahai, R., Bujarrabal, V., Castro-Carrizo, A., \& Zijlstra, A.  2000a, \aap, 360, L9
\bibitem[Sahai et al.(2007)]{sah07} Sahai, R., Morris, M., S\'{a}nchez Contreras, C., \& Claussen, M.  2007, \aj, 134, 2200
\bibitem[Sahai et al.(2000b)]{sah00}  Sahai, R., Su, K.Y.L., Kwok, S., Dayal, A., \& Hrivnak, B.J.  2000b, in ASP Conf. Ser. 199, Asymmetrical Planetary Nebulae II:  From Origins to Microstructures, ed. J.H.  Kastner, N. Soker, \& S. Rappaport (San Francisco:  ASP), 167
\bibitem[Sahai et al.(1991)]{sah91} Sahai, R., Wootten, A., Schwarz, H.E., \& Clegg, R.E.S.  1991,\aap, 251, 560
\bibitem[Sahai \& Trauger(1998)]{sah98} Sahai, R., \& Trauger, J.T.  1998, \aj, 116, 1357
\bibitem[Sahai et al.(1998)]{saa98} Sahai, R., Trauger, J.T., Watson, A.M., et al.  1998, \apj, 493, 301
\bibitem[Sevenster(2002)]{sev02}  Sevenster, M.N.  2002, \aj, 123, 2772
%\bibitem[Soker(2002)]{sok02} Soker, N.  2002, ApJ, 568, 726
\bibitem[Su\'arez et al.(2008)]{sua08}  Su\'arez, O., G\'omez, J.F., \& Miranda, L.F.  2008, \apj, 689, 430
\bibitem[Van de Steene \& Pottasch(1995)]{van95}  Van de Steene, G.C., \& Pottasch, S.R.  1995, \aap, 299, 238 
%\bibitem[Vlemmings et al.(2006)]{vle06} Vlemmings, W.H.T., Diamond, P.J., \& Imai, H.  2006, Nature, 440, 58
%\bibitem[Westbrook et al.(1975)]{wes75} Westbrook, W.E., Willner, S.P., Merrill, K.M., Schmidt, M., Becklin, E.E., Neugebauer, G., \& Wynn-Williams, C.G.  1975, ApJ, 202, 407
\end{thebibliography}
\end{document}